\documentclass[11pt]{article}

\usepackage[preprint]{acl}

\usepackage{times}
\usepackage{latexsym}
\usepackage[T1]{fontenc}
\usepackage[utf8]{inputenc}
\usepackage{microtype}
\usepackage{inconsolata}

\usepackage{graphicx}
\usepackage{natbib}
\usepackage{booktabs}

\usepackage{amsmath}
\usepackage{xcolor}   
\usepackage{arydshln} 
\usepackage{multirow} 

\usepackage{algorithm}
\usepackage{algorithmic}

\hypersetup{
    pdftitle={MERGE: Multi-LLM Ensemble for Retrieval via Generative Enrichment},
    pdfauthor={Tzu-I Ho, Yung-Yu Shih, Shang-Yu Su, Dongzhe Wang, Yun-Nung Chen},
    pdfkeywords={query expansion, information retrieval, large language models, ensemble, automatic prompt optimization, BM25, BEIR}
}

\title{MERGE: Multi-LLM Ensemble for Retrieval via Generative Enrichment}

\author{
    \textbf{Tzu-I Ho}\textsuperscript{1}\thanks{Equal contribution. Work done at Rakuten.},\;
    \textbf{Yung-Yu Shih}\textsuperscript{2}\footnotemark[\value{footnote}]\thanks{Work done during an internship at Taiwan Rakuten Ichiba, Inc.},\;
    \textbf{Shang-Yu Su\textsuperscript{3}},\;
    \textbf{Dongzhe Wang\textsuperscript{4}},\;
    \textbf{Yun-Nung Chen\textsuperscript{2}}
    \\
    \textsuperscript{1}University of Waterloo \quad
    \textsuperscript{2}National Taiwan University \\
    \textsuperscript{3}Rakuten Group, Inc.\ \quad
    \textsuperscript{4}Rakuten Asia Pte.\ Ltd. \\
    \texttt{tzuiho.tw@gmail.com, f12944007@ntu.edu.tw,} \\
    \texttt{shangyu.su@rakuten.com, dongzhe.wang@rakuten.com, y.v.chen@ieee.org}
}

\begin{document}

\maketitle

\begin{abstract}
Large Language Models (LLMs) are increasingly used to enrich user queries in information retrieval (IR) so that a standard retriever such as BM25 can bridge vocabulary gaps with the target corpus. Any single LLM, however, is limited by its training data and architectural biases, and its enrichment behavior depends on hand-crafted prompts that must be re-engineered for each new model---an expensive and poorly scalable process. We present \textbf{MERGE} (\emph{Multi-LLM Ensemble for Retrieval via Generative Enrichment}), a two-stage framework: three heterogeneous 7--8B open-source LLMs independently produce candidate expansions, and a larger LLM generatively synthesizes them into a single query. To make prompt engineering scalable across the ensemble, we integrate a \emph{task-grounded} Automatic Prompt Optimization (APO) loop into both stages. Unlike APO methods that judge candidates with an LLM evaluator, our loop scores each candidate by its downstream retrieval performance and runs a small tournament between the current champion prompt and optimizer-proposed drafts, terminating once the champion survives two consecutive rounds; a history-augmented variant additionally feeds the recent tournament trajectory back to the optimizer. MERGE is retriever-agnostic and issues a single BM25 pass with no rank fusion, no supervised document expansion, and no re-indexing. On five BEIR benchmarks (NQ, SciFact, FiQA, Touch\'e-2020, DBPedia), MERGE improves BM25 nDCG@10 over the original queries by $+2.1$ to $+14.9$ points and matches or outperforms strong LLM-based query-expansion baselines despite using only compact open-source models. Ablations confirm that the Stage-2 ensemble beats any single Stage-1 LLM, and that task-grounded APO converts large seed-prompt \emph{regressions} into consistent gains without hand-tuning.
\end{abstract}


\section{Introduction}

Large Language Models (LLMs) have become central components of modern information retrieval (IR) pipelines~\citep{zhu2023llmir}. For query-side enrichment in particular, methods such as Query2Doc~\citep{wang2023query2doc} and HyDE~\citep{gao2023hyde} use an LLM to generate pseudo-documents or hypothetical answers that clarify user intent and improve retrieval. The common thread is \emph{contextual query enrichment}: an LLM rewrites the user query so that a downstream retriever---often a strong lexical retriever such as BM25---can more easily match it with relevant passages.

Two problems limit the practical impact of this line of work. First, LLMs inherit biases and blind spots from their training data and architecture~\citep{bender2021dangers}, so a single model tends to produce enrichments with a distinctive but narrow style---missing rephrasings, hallucinating details, or over-committing to a specific interpretation of an ambiguous query. Second, controlling enrichment behavior relies heavily on hand-crafted prompts, and prompts that work well for one model family (e.g., Qwen2.5) transfer poorly to another (e.g., Llama-3.3 or Mistral) without a further round of manual tuning. This burden grows with the number of LLMs one wishes to combine, making multi-LLM ensembles hard to scale and hard to justify on cost grounds.

\begin{figure*}[t]
\centering
\includegraphics[width=0.92\textwidth]{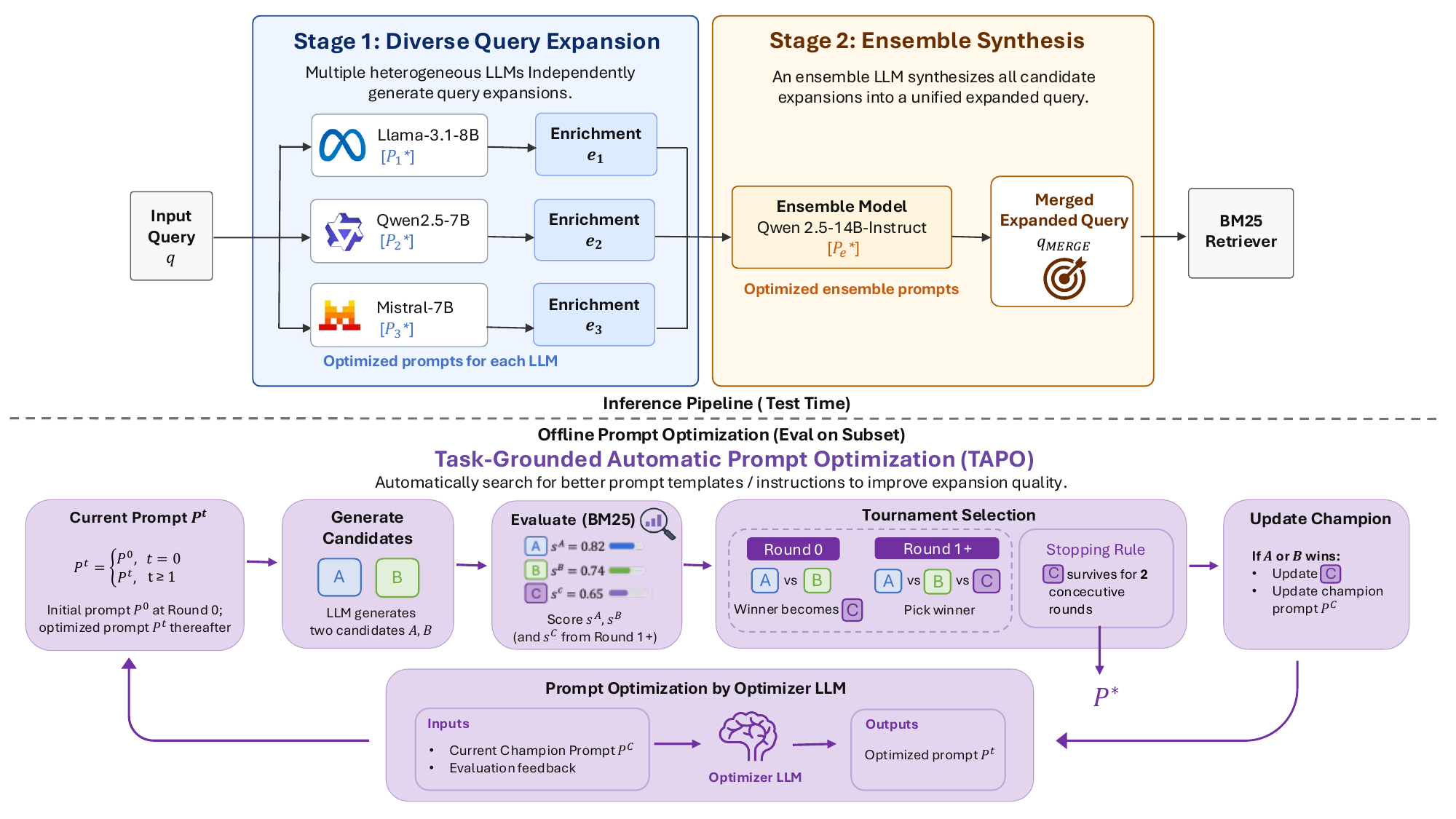}
\caption{Overview of MERGE. Given an input query, \textbf{Stage~1 (Diverse Query Expansion)} runs three heterogeneous 7--8B instruction-tuned LLMs (Llama-3.1-8B-Instruct, Qwen2.5-7B-Instruct, Mistral-7B-Instruct-v0.3) in parallel to produce candidate expansions $e_1, e_2, e_3$; \textbf{Stage~2 (Ensemble Synthesis)} uses Qwen2.5-14B-Instruct to generatively fuse the candidates into a single merged expanded query, which is then fed to a fixed BM25 retriever. Every prompt---each Stage-1 generator prompt and the Stage-2 ensemble prompt---is specialized via our \textbf{task-grounded Automatic Prompt Optimization} loop (bottom): starting from a hand-crafted seed, the optimizer LLM rewrites the current champion into two drafts, all three are scored on a 20\% dev subset via BM25 nDCG@10, and a tournament selects the winner as the next champion. The loop terminates when the champion survives two consecutive rounds.}
\label{fig:framework}
\end{figure*}

We address both problems with \textbf{MERGE} (\emph{Multi-LLM Ensemble for Retrieval via Generative Enrichment}). MERGE processes an input query $x$ through two stages, illustrated in Figure~\ref{fig:framework}:
\begin{itemize}
    \item \textbf{Generation.} Multiple heterogeneous LLMs independently produce enriched query candidates $y_1,\dots,y_n$, exposing diverse but complementary views of the user's intent. Each generator uses a prompt that has been automatically specialized to its own model.
    \item \textbf{Ensemble.} An LLM \emph{generatively} synthesizes $\{y_1,\dots,y_n\}$ into a single unified query expansion $\hat{y}$, filtering likely hallucinations and merging complementary information under an APO-tuned ensemble prompt.
\end{itemize}

MERGE embodies \emph{dual-level ensemble}: horizontal ensemble across heterogeneous generators within Stage~1, and vertical ensemble across the specialized Stage-1 / Stage-2 pipeline~\citep{dietterich2000ensemble}. Compared to selection-based fusion~\citep{jiang2023llmblender,tekin2024llmtopla} it retains full generative flexibility; compared to single-stage LLM expansion~\citep{wang2023query2doc,gao2023hyde} it explicitly separates \emph{producing} diverse candidates from \emph{synthesizing} them.

To address the scalability concern above, we integrate a \emph{task-grounded} Automatic Prompt Optimization (APO) loop into \emph{both} stages of MERGE. Starting from hand-crafted seed prompts, our APO loop iteratively rewrites and evolves each generator's prompt and the ensemble prompt under a retrieval-driven objective; each LLM ultimately runs with a prompt specialized to itself rather than a one-size-fits-all template. Unlike prior APO methods that rely on an LLM judge, our loop scores each candidate prompt by its actual retrieval performance on a small labelled subset and feeds this numerical signal, together with the history of previously tried prompts, back into the optimizer LLM. As a result, adding a new LLM to the generation stage no longer requires manual prompt engineering, and improvements in one LLM's prompt do not silently degrade the ensemble's behavior. Since APO is applied once and reused, its optimization cost is amortized over all downstream retrieval runs.

We evaluate MERGE on five diverse BEIR benchmarks~\citep{thakur2021beir}---NQ~\citep{kwiatkowski2019nq}, SciFact~\citep{wadden2020scifact}, FiQA~\citep{maia2018fiqa}, Touch\'e-2020~\citep{bondarenko2020touche}, and DBPedia~\citep{hasibi2017dbpedia}---covering open-domain QA, scientific claim verification, financial opinion QA, argument retrieval, and entity retrieval. Our contributions are as follows:
\begin{itemize}
    \item We propose MERGE, an IR-focused two-stage LLM ensemble that enriches user queries by combining heterogeneous multi-LLM generation with a generative ensemble step.
    \item We introduce a \emph{task-grounded} APO loop that scores candidate prompts by their actual downstream retrieval performance rather than by an LLM judge, and apply it to \emph{both} stages of MERGE. This turns per-LLM prompt engineering from a manual, non-scalable step into an automatic, model-specific optimization.
    \item MERGE is retriever-agnostic: its output is a plain-text query that can be fed to any retriever. Across five BEIR datasets, MERGE-enriched queries improve BM25 retrieval over the original queries on all five benchmarks and match or outperform strong LLM-based query-expansion baselines, with ablations that isolate the contributions of the ensemble stage and of APO. We adopt BM25 as the shared retrieval backbone in our experiments, matching the standard setup used in prior LLM-based query-expansion work, so that measured gains reflect \emph{how} the query is manipulated rather than a jointly trained retriever.
\end{itemize}

\section{Related Work}

\paragraph{LLM Ensembles and Output Fusion.}
Recent surveys~\citep{lu2024collaborative} categorize LLM ensembles into \emph{ensemble-before-inference} (routing~\citep{shnitzer2023routing,ong2024routellm}), \emph{during-inference} (token-level fusion~\citep{huang2024deepen,mavromatis2024packllm,xu2025sweetspan}), and \emph{after-inference} (response-level selection~\citep{li2024moreagents,guha2024smoothie} or regeneration~\citep{jiang2023llmblender,tekin2024llmtopla}). LLM-Blender~\citep{jiang2023llmblender} pioneered selection-then-regeneration, LLM-TOPLA~\citep{tekin2024llmtopla} maximizes response diversity in selection, and self-consistency~\citep{wang2023selfconsistency} majority-votes over reasoning paths. MERGE falls into the ensemble-after-inference family and is distinctive in (a) targeting IR enrichment specifically and (b) tuning every prompt in the ensemble automatically via task-grounded APO with a retrieval objective.

\paragraph{Query Expansion for IR.}
Classical query expansion uses relevance feedback~\citep{rocchio1971relevance} and embedding-based extensions~\citep{kuzi2016query}. Recent LLM-based methods---Query2Doc~\citep{wang2023query2doc}, HyDE~\citep{gao2023hyde}, and Exp4Fuse~\citep{liu2025exp4fuse}---use a single LLM to rewrite the query, with Exp4Fuse additionally fusing two BM25 ranked lists; we use these as our main baselines. doc2query~\citep{nogueira2019doc2query,gospodinov2023doc2query} and docT5query~\citep{nogueira2019doctttttquery} target the same query--corpus vocabulary gap but use a \emph{supervised} T5 fine-tuned on MS MARCO to generate query-like text \emph{from each document}, requiring corpus access and re-indexing when the corpus changes. Closely related is BEATS~\citep{shih2026beats}, a multi-stage LLM pipeline that enriches the \emph{corpus} side of e-commerce search through a human-in-the-loop prompt-refinement cycle. MERGE targets the same vocabulary gap without supervision or corpus access---operating on the user query only with zero-shot LLMs---and removes BEATS's human step by grounding prompt refinement in downstream retrieval scores.

\paragraph{Automatic Prompt Optimization.}
APO methods treat prompts as optimizable objects: ProTeGi~\citep{pryzant2023protegi} uses LLM-produced ``textual gradients''; APE~\citep{zhou2023ape} and OPRO~\citep{yang2024opro} cast prompt design as black-box optimization; evolutionary variants EvoPrompt~\citep{guo2024evoprompt} and Promptbreeder~\citep{fernando2023promptbreeder} maintain a prompt population; SPO~\citep{xiang2025spo} removes label supervision by using an LLM evaluator over pairs of outputs. Recent surveys~\citep{ramnath2025apo_survey} cover the space. Our loop builds on SPO but replaces the LLM evaluator with the actual downstream IR score on a small labelled subset, and feeds that numerical signal---together with the history of previously tried prompts---back into the optimizer LLM (\S\ref{sec:apo}).

\section{Methodology}

\subsection{Framework Overview}
Given an input query $x$, MERGE produces an enriched query $\hat{y}$ via
\begin{equation}
\hat{y} \;=\; M_e\!\left(P_e^*,\; \big\{\,M_i(P_i^*, x)\,\big\}_{i=1}^{n}\right),
\label{eq:merge}
\end{equation}
where $M_1,\dots,M_n$ are the $n$ heterogeneous Stage-1 LLMs, $M_e$ is the ensemble LLM at Stage~2, and each starred prompt $P_i^*$ (Stage 1) or $P_e^*$ (Stage 2) is produced by APO from a hand-crafted seed. Figure~\ref{fig:framework} illustrates the pipeline. The enriched query $\hat{y}$ is passed as-is to a BM25 retriever, and the passage corpus is used unmodified. Because $\hat{y}$ is plain text, MERGE is in principle retriever-agnostic and can also feed a dense retriever; we focus on BM25 in this work to match the standard evaluation protocol for LLM-based query expansion.

\subsection{Stage 1: Multi-LLM Generation}
We employ $n$ diverse LLMs $\{M_1,\dots,M_n\}$ selected to maximize heterogeneity in training data, architecture, and scale. Each generator receives its own APO-optimized prompt $P_i^*$ and the input query $x$, and produces a candidate $y_i = M_i(P_i^*, x)$. The seed prompt instructs the model to clarify user intent, add plausible synonyms and related concepts, and produce a refined search string aimed at improving recall. Terms that could leak the answer directly---rather than help locate a relevant passage---are discouraged.

\subsection{Stage 2: Generative Ensemble}
An ensemble model $M_e$ generatively synthesizes $\{y_1,\dots,y_n\}$ into a single enriched query $\hat{y}$. The APO-optimized ensemble prompt $P_e^*$ instructs $M_e$ to (i) identify the core search intent and high-value entities across candidates, (ii) resolve semantic redundancies and prioritize terms strongly aligned with the query, and (iii) synthesize a compact query expansion inside explicit \texttt{<result>} tags. We keep $M_e$ strictly distinct from every Stage-1 model to avoid self-preference bias in the fusion step.

\subsection{Task-Grounded Automatic Prompt Optimization}
\label{sec:apo}

Hand-crafting prompts for each LLM at each stage does not scale as one adds more models or moves to new model families. We remove this bottleneck by wrapping every (stage, model) pair with a task-grounded APO loop, applied to (a) each generator's prompt $P_i$ in Stage~1 and (b) the ensemble prompt $P_e$ in Stage~2.

\paragraph{Starting point: Self-Supervised Prompt Optimization.}
We build on Self-Supervised Prompt Optimization (SPO)~\citep{xiang2025spo}, a recent APO framework in which an \emph{optimizer LLM} iteratively rewrites a prompt while an \emph{evaluator LLM} decides which prompt to keep by comparing pairs of outputs. SPO does not require ground-truth labels because its evaluator is itself an LLM judge. This makes SPO attractive as a starting point but leaves open a well-known concern for ensemble settings: an LLM judge introduces its own biases and its evaluation signal is only loosely coupled to the ultimate downstream metric one cares about.

\paragraph{Our modification: task-grounded evaluation.}
For MERGE, the downstream metric we care about is retrieval quality, and it is cheap to measure. We therefore replace SPO's LLM evaluator with the actual downstream IR task: each candidate prompt is scored by running the corresponding stage end-to-end on a fixed \emph{20\% subset} of the training data and measuring nDCG@10 with the same fixed BM25 retriever used at test time. This gives the optimizer a task-grounded numerical signal rather than the pairwise LLM-judged preference used in vanilla SPO.

\paragraph{Tournament-style optimization loop.}
Rather than accepting or rejecting a single new prompt per round, each round of our APO loop is framed as a small tournament (Algorithm~\ref{alg:apo}). Two LLM roles are involved: a \emph{generator} $G$ that produces query-expansion sets from a prompt, and an \emph{optimizer} $O$ that rewrites prompts. The tournament tracks tuples of the form (expansion set, retrieval score, prompt-of-origin). In round~0, $G$ samples two expansion sets $A, B \sim G(P^0)$ from the hand-crafted seed prompt $P^0$, each is scored on the evaluation subset via the fixed retriever $R$, and the higher-scoring tuple becomes the initial champion $(C, s_C, P^C)$ with $P^C = P^0$. From round $t \geq 1$ onwards, $O$ first rewrites the champion prompt into a new draft $P^t = O(P^C, \text{evidence}_t)$, then $G$ samples two expansion sets $A, B \sim G(P^t)$ from $P^t$, and the highest-scoring tuple among the two challengers and the incumbent champion becomes the new champion. When a challenger wins, the champion tuple is replaced entirely (expansion, score, and prompt). Drawing \emph{two} expansions per prompt rather than one exposes more of what a prompt can produce under stochastic decoding, guards against early convergence to a champion whose lead came from a single lucky decode, and forces challengers to clear a stricter bar before they can unseat the incumbent. We treat the loop as \emph{converged} when the champion survives two consecutive rounds without being beaten, signalling that the optimizer has run out of profitable edits, and return $P^* = P^C$.

\paragraph{What the optimizer sees each round.}
At each round $O$ rewrites the champion prompt $P^C$ into a new draft prompt $P^t$ under a structured optimizer prompt that supplies (a) a numerical signal comparing the previous round's challenger expansions to the champion's expansion (score difference, 95\% CI, p-value on the evaluation subset), (b) automatically extracted \emph{wins} and \emph{losses} examples where a challenger's expansion changed retrieval outcome relative to the champion, and (c) \emph{theme signals} such as gold-aligned terms the challengers are missing and noise terms they introduce. $O$ is explicitly instructed to make a substantive edit that addresses observed regressions while preserving working strategies. We consider two variants of this optimizer prompt that differ only in whether $O$ also sees a short \emph{tournament history} of the previous rounds (winner, expansions produced, and BM25 scores):
\begin{itemize}
    \item \textbf{MERGE (APO)}, our default: the optimizer sees only the current-round signals above.
    \item \textbf{MERGE (APO w/ history)}: the optimizer additionally sees the last five rounds so that it can condition its rewrite on \emph{both} the textual trajectory of past prompts and their measured scores.
\end{itemize}

APO is applied stage-wise: to every Stage-1 generator first, then to the Stage-2 ensemble prompt with those Stage-1 prompts frozen.

\begin{algorithm}[t]
\caption{Task-grounded APO for one (stage, model) pair.}
\label{alg:apo}
\textbf{Input}: seed prompt $P^0$, generator $G$, optimizer $O$, retriever $R$, eval subset $\mathcal{D}_e$, max rounds $T$ \\
\textbf{Output}: optimized prompt $P^*$
\begin{algorithmic}[1]
\STATE \emph{// Round 0: two expansion samples from the seed prompt}
\STATE $(A, B) \gets G(P^0)$ \hfill \emph{// two sampled expansions}
\STATE $s_A \gets \text{IR}(A, R, \mathcal{D}_e)$;\; $s_B \gets \text{IR}(B, R, \mathcal{D}_e)$
\STATE $(C, s_C, P^C) \gets$ higher of $\{(A, s_A, P^0),\,(B, s_B, P^0)\}$
\STATE $\text{streak} \gets 0$
\FOR{$t = 1$ to $T$}
    \STATE $P^t \gets O(P^C, \text{evidence}_t)$ \hfill \emph{// evolve prompt}
    \STATE $(A, B) \gets G(P^t)$ \hfill \emph{// two new expansions from $P^t$}
    \STATE $s_A \gets \text{IR}(A, R, \mathcal{D}_e)$;\; $s_B \gets \text{IR}(B, R, \mathcal{D}_e)$
    \STATE $(W, s_W, P^W) \gets \arg\max \{(A, s_A, P^t),$ \\
    \hspace*{4.9em}$(B, s_B, P^t),\,(C, s_C, P^C)\}$
    \IF{$P^W \equiv P^C$}
        \STATE $\text{streak} \gets \text{streak} + 1$
        \IF{$\text{streak} \geq 2$}
            \STATE \textbf{break} \hfill \emph{// champion survived twice}
        \ENDIF
    \ELSE
        \STATE $(C, s_C, P^C) \gets (W, s_W, P^W)$;\; $\text{streak} \gets 0$
    \ENDIF
\ENDFOR
\STATE \textbf{return} $P^* \gets P^C$
\end{algorithmic}
\end{algorithm}

\paragraph{Why task-grounded APO helps MERGE.} Our APO loop delivers three benefits that are essential for a multi-LLM ensemble like MERGE. \emph{(1) Scalability across LLMs.} Adding a new generator no longer requires a human to hand-tune its prompt; the tournament produces a model-specific prompt automatically. \emph{(2) Cost.} The dominant cost of MERGE at inference time is running many LLMs; hand-tuning prompts on top of this is a hidden but substantial developer cost. APO amortizes this cost into a one-off optimization phase whose budget the user can control. \emph{(3) Consistency.} Because every prompt is optimized against the same retrieval-driven objective, prompts across the ensemble are tuned against a common yardstick, which makes the Stage-2 ensemble output more predictable.

\subsection{Dual-Level Ensemble}

MERGE operates ensemble at two complementary levels~\citep{dietterich2000ensemble}. \emph{Horizontally}, Stage~1 aggregates heterogeneous LLMs that differ in architecture, training data, and scale---analogous to bagging, where diversity among base learners reduces variance and mitigates single-model blind spots. \emph{Vertically}, Stage~2 specializes in synthesis (filtering hallucinations and merging complementary information into one compact expansion)---analogous to stacking, where a meta-model integrates base predictions. APO acts orthogonally to both levels, tuning \emph{how} each model plays its role without changing the overall dual-level structure.

\section{Experiments}

\subsection{Setup}

\paragraph{Datasets and metric.} We evaluate on five heterogeneous BEIR~\citep{thakur2021beir} benchmarks spanning open-domain QA (NQ~\citep{kwiatkowski2019nq}), scientific-claim retrieval (SciFact~\citep{wadden2020scifact}), financial opinion retrieval (FiQA~\citep{maia2018fiqa}), argument retrieval (Touch\'e-2020~\citep{bondarenko2020touche}), and entity retrieval (DBPedia~\citep{hasibi2017dbpedia}). This suite covers markedly different domains, query styles, and passage-length regimes. MERGE (with APO-tuned prompts) is applied to all queries; each passage corpus is left unmodified. Following BEIR conventions we report nDCG@10.

\paragraph{Models and retriever.}
Stage~1 uses three 7--8B instruction-tuned models from three different families---\textbf{Qwen2.5-7B-Instruct}~\citep{qwen2024technical}, \textbf{Meta-Llama-3.1-8B-Instruct}~\citep{llama3modelcard,touvron2023llama}, and \textbf{Mistral-7B-Instruct-v0.3}~\citep{jiang2023mistral}---keeping the pool at a comparable scale so no single generator dominates by size while different pretraining regimes provide the horizontal diversity Stage~2 exploits. Stage~2 uses \textbf{Qwen2.5-14B-Instruct}~\citep{qwen2024technical}, which is also the APO optimizer. All methods share a fixed \textbf{BM25}~\citep{robertson2009bm25} backbone, matching the standard setup for LLM-based query expansion~\citep{wang2023query2doc,gao2023hyde} so that measured gains reflect how the query is manipulated rather than a jointly trained retriever. LLM inference uses vLLM~\citep{kwon2023vllm} on NVIDIA H100 GPUs.

\paragraph{Baselines.}
All methods share the same BM25 retriever; they differ only in how LLM-generated text is used to bridge the query--corpus vocabulary gap. We compare MERGE against three published LLM-based expansion baselines and a no-APO ablation:
(i) \textbf{Vanilla BM25}: the original query with no expansion, the reference point;
(ii) \textbf{docT5query}~\citep{nogueira2019doctttttquery}: a \emph{supervised, corpus-dependent} baseline that fine-tunes T5 on MS MARCO (document, query) pairs so it can predict queries from documents, then runs T5 on every document in the corpus and appends the predicted queries to that document in the BM25 index---the LLM's input is documents;
(iii) \textbf{BM25+Exp4Fuse}~\citep{liu2025exp4fuse}: an unsupervised, query-only baseline that runs two BM25 passes (over the original query and over an LLM-produced pseudo-document generated from the user query) and combines the two ranked lists with a modified reciprocal-rank fusion;
(iv) \textbf{docT5query+Exp4Fuse}~\citep{liu2025exp4fuse,nogueira2019doctttttquery}: the stronger baseline reported by the Exp4Fuse paper, stacking (ii) and (iii);
(v) \textbf{MERGE w/o APO}: MERGE with the same two-stage architecture but using only the hand-crafted seed prompts, to isolate the effect of prompt optimization.
Like Exp4Fuse variants and unlike docT5query, MERGE takes only the user query as input to its LLMs; unlike Exp4Fuse variants, MERGE issues a single BM25 pass with no rank fusion.

For MERGE with APO, we report both variants introduced in \S\ref{sec:apo}: \textbf{MERGE (APO)}, the default tournament optimizer, and \textbf{MERGE (APO w/ history)}, the history-augmented variant.

\paragraph{Reproducibility.}
All experiments rely on publicly available BEIR benchmarks and open-source model checkpoints (Qwen2.5-7B/14B-Instruct, Llama-3.1-8B-Instruct, Mistral-7B-Instruct-v0.3) with a fixed BM25 backbone. The APO tournament (Algorithm~\ref{alg:apo}), the two APO variants (\S\ref{sec:apo}), the seed-prompt design (\S\ref{sec:apo}), and the 20\% dev-subset scoring protocol are fully described in the paper. Because MERGE requires only these public inputs and a lexical retriever with standard hyperparameters---no supervised training data, no proprietary corpus, no learned ranker---the results in Tables~\ref{tab:main} and~\ref{tab:ablation} can be independently reimplemented from the descriptions given here. Following widely used BM25 practice, we repeat the original query several times and append the enrichment as the retrieval input.

\subsection{Main Results}

Table~\ref{tab:main} reports nDCG@10 on the five BEIR benchmarks. All rows share the same BM25 retriever and differ only in how the query is manipulated before it is fed to BM25. Numbers for BM25+Exp4Fuse and T5+Exp4Fuse are taken from~\citet{liu2025exp4fuse} under the same BM25 evaluation protocol; MERGE variants use the models described above with our task-grounded APO applied to every prompt.

\begin{table*}[t]
\caption{Retrieval results on five BEIR benchmarks (nDCG@10, \%). All rows share the same BM25 retriever and differ only in how the query is expanded. Numbers in parentheses are reproduced from~\citet{liu2025exp4fuse}; docT5query is~\citet{nogueira2019doctttttquery}. \textbf{Bold} = best result in each column; \underline{underline} = second best.}
\label{tab:main}
\centering
\footnotesize
\setlength{\tabcolsep}{6pt}
\begin{tabular}{lccccc}
\toprule
Method & NQ & SciFact & FiQA & Touch\'e-2020 & DBPedia \\
\midrule
Vanilla BM25 (original query)                & 30.48 & 67.62 & 23.85 & 44.23 & 31.99 \\
\; + docT5query                              & (38.10) & (67.50) & (25.20) & (34.70) & (33.10) \\
\; + BM25 + Exp4Fuse                         & (39.10) & (68.80) & (24.70) & (51.20) & (36.10) \\
\; + docT5query + Exp4Fuse                   & (42.80) & (\underline{71.30}) & (\textbf{26.30}) & (39.90) & (\textbf{38.90}) \\
\hdashline
\; + MERGE w/o APO                           & 34.52 & 67.79 & 19.41 & 38.69 & 33.59 \\
\; \textbf{+ MERGE (APO)}                    & \textbf{45.42} & \textbf{71.40} & 25.59 & \underline{50.16} & \underline{37.94} \\
\; \textbf{+ MERGE (APO w/ history)}         & \underline{42.50} & 70.01 & \underline{25.91} & \textbf{51.67} & 37.84 \\
\bottomrule
\end{tabular}
\end{table*}

\paragraph{MERGE vs.\ vanilla BM25.}
Across all five datasets, MERGE with APO improves BM25 nDCG@10 over the \emph{no-expansion} reference. The absolute gain is largest where the original queries are furthest from the corpus vocabulary or style: $+14.9$ points on NQ (from $30.48$ to $45.42$), $+7.4$ on Touch\'e-2020, $+6.0$ on DBPedia, and $+3.8$ on SciFact. On FiQA the gain is smaller but still positive ($+2.1$ points, $23.85\!\to\!25.91$). This is consistent with MERGE closing a vocabulary/style gap on the query side, exactly where a lexical retriever like BM25 is most vulnerable.

\paragraph{MERGE vs.\ LLM-expansion baselines.}
Compared with docT5query alone---a supervised, corpus-dependent baseline that runs T5 on every document---MERGE wins on all five benchmarks: $+7.3$ on NQ, $+3.9$ on SciFact, $+16.9$ on Touch\'e-2020, and $+4.8$ on DBPedia; on FiQA the two are close (MERGE APO w/ history 25.91 vs.\ docT5query 25.20, a $+0.7$ win). Compared with BM25+Exp4Fuse, MERGE (in either APO variant) is better on all five datasets, with gains from $+0.5$ (Touch\'e-2020) to $+6.3$ (NQ) points. Compared with the stronger stacked baseline docT5query+Exp4Fuse, MERGE wins clearly on NQ ($+2.6$) and Touch\'e-2020 ($+11.8$), is essentially tied on SciFact ($+0.1$), and loses narrowly on FiQA ($-0.4$) and DBPedia ($-1.0$). This is a favourable outcome: MERGE beats every unsupervised LLM-based query-expansion baseline on 5/5 benchmarks and matches or exceeds the strongest supervised baseline on 3/5, using only compact 7--8B open-source models, no supervised query--document pairs, and no access to the corpus.

\paragraph{Simplicity, cost, and productionizability.}
Beyond raw retrieval numbers, MERGE is structurally simpler than the alternatives in Table~\ref{tab:main}: docT5query is \emph{supervised and corpus-dependent} (needs a labelled MS MARCO training set, runs T5 on every document, requires re-indexing when the corpus changes), and the Exp4Fuse variants add two BM25 passes and a reciprocal-rank-fusion step with tunable weights~\citep{liu2025exp4fuse}. MERGE has none of these dependencies: its LLMs see only the user query, are used zero-shot, and produce an enriched query $\hat{y}$ that is issued to a single BM25 pass over the original corpus. The only extra cost is the one-off APO tournament (\S\ref{sec:apo}), which typically converges in under 10 rounds per (stage, model) pair on a 20\% dev subset (well under one hour on a single H100) and is amortized over all downstream retrieval traffic. This substitutes a small compute cost for the per-LLM manual prompt-engineering effort that would otherwise dominate as the model pool grows; empirically (Table~\ref{tab:main}, M0 vs.\ M1/M2), that compute cost translates directly into large retrieval gains. Being corpus-agnostic also means MERGE can be applied at query time for online retrieval or offline as training-data augmentation for a downstream dense retriever, without changes when the corpus changes.

\paragraph{Effect of APO (M0 vs.\ M1/M2).} Comparing MERGE~w/o~APO (M0) against the two APO variants isolates the effect of prompt optimization. Without APO, seed prompts \emph{hurt} retrieval on two out of five datasets: on FiQA ($-4.4$ vs.\ vanilla BM25) and Touch\'e-2020 ($-5.5$)\%  ; a mis-specified prompt can actively degrade BM25 by adding noisy terms. Adding APO turns those regressions into consistent gains across all five datasets, without requiring any additional manual engineering. Between M0 and the better APO variant we see gains of $+10.9$ (NQ), $+3.6$ (SciFact), $+6.5$ (FiQA), $+13.0$ (Touch\'e-2020), and $+4.4$ (DBPedia) nDCG@10 points. This is important not only because the absolute gain is large but because it converts prompt tuning from a manual per-LLM step into an automatic one.

\paragraph{APO w/o vs.\ w/ history.}
The two APO variants trade off in a query-dependent way. \emph{MERGE (APO)} is stronger on precise factoid/entity queries (NQ $45.42$ vs.\ $42.50$; SciFact $71.40$ vs.\ $70.01$; DBPedia $37.94$ vs.\ $37.84$), while \emph{MERGE (APO w/ history)} is stronger on broader, multifaceted queries (Touch\'e-2020 $51.67$ vs.\ $50.16$; FiQA $25.91$ vs.\ $25.59$). We hypothesize that feeding the last five rounds of the tournament back to the optimizer helps in domains where useful expansions require exploring a wider space of phrasings (financial and argumentative queries), while on entity-centric queries the extra history noise can distract from the direct win/loss signal.

\subsection{Ablation: Both Stages Contribute}
\label{sec:ablation}

Table~\ref{tab:ablation} isolates the contribution of the Stage-2 ensemble by comparing every individual Stage-1 generator's enriched query against the final MERGE output, on three representative benchmarks (NQ, SciFact, FiQA). To keep the comparison apples-to-apples, all rows use the default APO variant (\emph{MERGE (APO)}) and the same BM25 retriever; the only thing that varies is whether the query fed to BM25 is a single Stage-1 model's expansion or the Stage-2 ensemble of all three.

\begin{table}[t]
\caption{Ablation on the effect of the two-stage architecture. nDCG@10 (\%) with BM25 backbone; all methods use \emph{MERGE (APO)} (M1). Rows 1--3: BM25 fed a single Stage-1 model's expansion. Row 4: BM25 fed the Stage-2 ensemble expansion.}
\label{tab:ablation}
\centering
\setlength{\tabcolsep}{4pt}
\small
\begin{tabular}{lccc}
\toprule
Query fed to BM25 & NQ & SciFact & FiQA \\
\midrule
Original query (no expansion)     & 30.48 & 67.62 & 23.85 \\
\midrule
Stage-1: Mistral-7B-v0.3          & 38.79 & 70.49 & 24.95 \\
Stage-1: Qwen2.5-7B               & 34.59 & 68.69 & 24.17 \\
Stage-1: Llama-3.1-8B             & 41.06 & 69.72 & 25.00 \\
\midrule
\textbf{Stage-2: Qwen2.5-14B (ours)} & \textbf{45.42} & \textbf{71.40} & \textbf{25.59} \\
\bottomrule
\end{tabular}
\end{table}

\begin{table*}[!t]
\caption{APO effect at Stage~1. Input query (NQ): \emph{``the cuban revolt against spain was led by''}. Model: Llama-3.1-8B-Instruct. We show the enriched query produced under the three APO regimes and its BM25 nDCG@10 on the NQ evaluation split. M0 uses the hand-crafted seed prompt; M1 and M2 use tournament-optimized prompts without / with a 5-round history.}
\label{tab:qual_apo}
\centering
\small
\setlength{\tabcolsep}{4pt}
\renewcommand{\arraystretch}{1.15}
\begin{tabular}{|c|c|p{0.65\textwidth}|}
\hline
\textbf{Regime} & \textbf{nDCG@10} & \textbf{Stage-1 output (Llama-3.1-8B-Instruct)} \\
\hline
M0 (no APO) & 33.31 & ``Jos\'e Mart\'i, Calixto Garc\'ia, Antonio Maceo, Carlos Manuel de C\'espedes\ldots'' (short entity list, no context) \\
\hline
M1 (APO)    & 41.06 & ``The Cuban revolt against Spain was led by Jos\'e Mart\'i, a Cuban poet and revolutionary. Mart\'i was a key figure in the \emph{Cuban War of Independence}, which began in \emph{1895} and lasted until \emph{1898}. He advocated for Cuban independence and was a critic of Spanish colonial rule\ldots'' \\
\hline
M2 (APO w/ history) & 40.14 & ``The Cuban revolt against Spain was led by Jos\'e Mart\'i, a Cuban poet, writer, and revolutionary. Mart\'i was a key figure in the Cuban independence movement and played a significant role in organizing the \emph{Cuban Revolutionary Party}, which aimed to overthrow \emph{Spanish rule} in Cuba\ldots'' \\
\hline
\end{tabular}
\end{table*}

\begin{table*}[!t]
\caption{Ensemble effect at Stage~2 on the NQ query \emph{``the cuban revolt against spain was led by''}. We show the three Stage-1 candidates (one per LLM in the pool) and the Stage-2 ensemble output, all under MERGE~(APO). ``Score'' is BM25 nDCG@10 on the NQ evaluation split. Long enrichments are elided as ``\ldots'' but no substantive content is added.}
\label{tab:qual_ensemble}
\centering
\small
\setlength{\tabcolsep}{4pt}
\renewcommand{\arraystretch}{1.15}
\begin{tabular}{|l|c|p{0.68\textwidth}|}
\hline
\textbf{Source} & \textbf{Score} & \textbf{Enriched output} \\
\hline
Stage-1: Mistral-7B    & 38.79 & ``The Cuban Revolution\ldots was led by a group of individuals who sought to overthrow the \emph{Spanish colonial rule} and establish an independent Cuban government\ldots armed conflicts from \emph{1868} to \emph{1898}\ldots'' \\
\hline
Stage-1: Qwen2.5-7B    & 34.59 & ``\ldots led by leaders, figures, insurgents, and activists who fought for, spearheaded, and initiated the\ldots'' (generic; no named entities) \\
\hline
Stage-1: Llama-3.1-8B  & 41.06 & ``\ldots led by \emph{Jos\'e Mart\'i}, a Cuban poet and revolutionary\ldots the \emph{Cuban War of Independence}, which began in \emph{1895} and lasted until \emph{1898}\ldots'' \\
\hline
\textbf{Stage-2: Qwen2.5-14B} & \textbf{45.42} & ``\ldots led by a group of Cuban revolutionaries, including \emph{Jos\'e Mart\'i}, \emph{M\'aximo G\'omez}, and \emph{Antonio Maceo}. The revolt began in \emph{1868} and lasted until \emph{1898}\ldots Mart\'i as the ideological leader and G\'omez and Maceo as military leaders\ldots'' \\
\hline
\end{tabular}
\end{table*}

Two observations. First, all three Stage-1 models with APO already improve BM25 over the original query, but the gain varies noticeably across models (e.g.\ on NQ, Llama beats Qwen by more than $6$ nDCG@10 points). Second, the Stage-2 ensemble strictly dominates the best individual Stage-1 model on all three datasets, adding $+4.4$ (NQ), $+0.9$ (SciFact), and $+0.6$ (FiQA) nDCG@10 on top of the best Stage-1 candidate. This confirms that both stages contribute to retrieval quality: Stage~1 provides diverse candidates, and Stage~2 extracts more than any single generator does alone. The MERGE~w/o~APO row in the main table complements this ablation by isolating the effect of the APO loop itself.

\subsection{Qualitative Analysis}

We illustrate the two levers introduced by MERGE with real enrichments from our runs. Table~\ref{tab:qual_apo} (\emph{APO effect}) fixes a single query and a single Stage-1 model and shows how the enrichment evolves as we move from no APO (M0) through the two APO variants (M1, M2). Table~\ref{tab:qual_ensemble} (\emph{Ensemble effect}) fixes MERGE~(APO) and shows, for two representative queries, all three Stage-1 candidates and the Stage-2 ensemble output. Long enrichments are elided as ``\ldots'' but no substantive content is added.

\paragraph{Observations.}
Table~\ref{tab:qual_apo} illustrates why APO helps: the seed prompt (M0) produces a bare entity list that loses BM25 signal, while tournament-optimized prompts (M1, M2) pair the same entities with retrieval-friendly context (movements, dates)---exactly what BM25 needs to match relevant Wikipedia passages. Table~\ref{tab:qual_ensemble} makes the horizontal diversity across Stage-1 models visible: Mistral supplies date ranges (\emph{1868--1898}), Llama supplies the key entity (\emph{Jos\'e Mart\'i}) with biography, and Qwen contributes almost nothing useful (score $34.59$). The Stage-2 ensemble fuses the useful signals (\emph{Mart\'i}, \emph{G\'omez}, \emph{Maceo}, \emph{1868--1898}, ideological vs.\ military leadership) into a single enrichment and reaches $45.42$, outperforming every individual Stage-1 output. These patterns align with Table~\ref{tab:main}: APO without history wins on precise entity queries (NQ) by preserving entity focus, while APO w/ history wins on exploratory queries (Touch\'e-2020, FiQA) by broadening the expansion space.

Two properties of the APO loop itself are worth flagging. \emph{Convergence.} Most tournaments hit the two-consecutive-wins termination in fewer than 10 rounds across models and datasets, which is what keeps the one-off APO cost small in practice. \emph{Optimizer-discovered prompts.} For some Stage-1 models the converged prompt is substantially longer than the seed and, while composed of real English words, reads more as a bag of model-specific steering tokens than as a natural-language instruction---a common by-product of optimizing prompts directly against a downstream metric rather than against human readability. Importantly, the query enrichments those prompts produce remain readable (Tables~\ref{tab:qual_apo}--\ref{tab:qual_ensemble}), so this does not affect end-to-end usability.

\section{Conclusion}

We presented MERGE, a two-stage LLM-ensemble framework for query enrichment that couples three heterogeneous 7--8B generators with a 14B generative ensemble step, and integrated a task-grounded Automatic Prompt Optimization (APO) loop into both stages: each prompt is scored by BM25 nDCG@10 on a 20\% dev subset, and a small tournament between the current champion and two optimizer-proposed drafts terminates when the champion survives two consecutive rounds; a history-augmented variant additionally feeds the last five rounds back to the optimizer. On five BEIR benchmarks with a shared BM25 backbone, MERGE improves nDCG@10 over the original queries on all five datasets (up to $+14.9$ on NQ), outperforms BM25+Exp4Fuse on every benchmark, and matches or exceeds the stronger docT5query+Exp4Fuse baseline on 3/5---using only compact open-source models and no supervised query--document pairs. Ablations show that the Stage-2 ensemble strictly dominates the best Stage-1 model and that task-grounded APO turns $-4$ to $-5$ point seed-prompt regressions into consistent gains without manual tuning, converting per-LLM prompt engineering from a manual step into an automatic one. Because MERGE is corpus-agnostic, $\hat{y}$ can also feed a dense retriever (at inference or as training-data augmentation), and the same two-stage ensemble could be extended to passage-side enrichment as in doc2query~\citep{nogueira2019doc2query} or BEATS~\citep{shih2026beats}.

\clearpage
\bibliography{references}

\end{document}